\input harvmac
\def\np#1#2#3{Nucl. Phys. B {#1} (#2) #3}
\def\pl#1#2#3{Phys. Lett. B {#1} (#2) #3}
\def\plb#1#2#3{Phys. Lett. B {#1} (#2) #3}

\def\physrev#1#2#3{Phys. Rev. D {#1} (#2) #3}

\nref\jptasi{J. Polchinksi, ``TASI Lectures on D-Branes,''
hep-th/9611050.} 
\nref\wsmall{E. Witten, ``Small Instantons in String Theory,''
hep-th/9511030, \np{460}{1995}{541.}}
\nref\gh{O. Ganor and A. Hanany, ``Small E(8) Instantons and
Tensionless Noncritical Strings,'' hep-th/9602120.}
\nref\sw{N. Seiberg and E. Witten, ``Comments on String Dynamics in
Six Dimensions,'' hep-th/9603003, \np{471}{1996}{121}.}
\nref\dlpt{M.J. Duff, H. Lu, and C.N. Pope, ``Heterotic Phase
Transitions and Singularities of the Gauge Dyonic String,''
hep-th/9603037, \plb{378}{1996}{101}.}
\nref\wsd{E. Witten, ``Physical Interpretation of Certain Strong
Coupling Singularities,'' hep-th/9609159, Mod. Phys. Lett. A11 (1996)
2649.}
\nref\Wcomm{E. Witten, ``Some Comments on String Dynamics,'' 
hep-th/9507121, Proc. of Strings '95, editors I. Bars et. al., World
Scientific, 1996.}
\nref\ismir{K. Intriligator and N. Seiberg, ``Mirror Symmetry in Three
Dimensional Gauge Theories,'' hep-th/9607207, \pl{387}{1996}{513}.}
\nref\sfdfp{N. Seiberg, ``Five Dimensional SUSY Field Theories,
Non-trivial Fixed Points, and String Dynamics,'' hep-th/9608111.}
\nref\gms{O. J. Ganor, D. R. Morrison, and N. Seiberg, ``Branes,
Calabi-Yau Spaces, and Toroidal Compactification of the $N=1$ Six
Dimensional $E_8$ Theory,'' hep-th/9610251.}
\lref\kron{P.B. Kronheimer, Jour. Differential Geometry, 
{\bf 29} (1989) 665.}
\lref\kn{P. B. Kronheimer and H. Nakajima, ``Yang-Mills Instantons on
ALE Gravitational Instantons,'' Math. Ann. 288 (1990) 263.}
\lref\dm{M. Douglas and G. Moore, ``D-Branes, Quivers, and ALE
Instantons,'' hep-th/9603167.}
\lref\sag{A. Sagnotti, ``A Note on the Green-Schwarz Mechanism in Open
String Theory,'' hep-th/9210127, \pl{294}{1992}{196}.}
\lref\lagw{L. Alvarez-Gaume' and E. Witten, ``Gravitational
Anomalies,'' \np{234}{1983}{269}.}
\lref\dmw{M. Duff, R.  Minasian, and E. Witten, ``Evidence for
Heterotic/Heterotic Duality,'' hep-th/9601036, \np{465}{1996}{413}.}
\lref\GP{E. G. Gimon and Polchinski, ``Consistency Conditions for
Orientifolds and D-Manifolds,'' hep-th/9601038,
\physrev{54}{1996}{1667}.}
\lref\GJ{E. G. Gimon and C. V. Johnson, ``K3 Orientifolds,''
hep-th/9604129, \np{477}{1996}{715}.}
\lref\DP{A. Dabholkar and J. Park, ``Strings on Orientifolds,''
hep-th/9604178, \np{477}{1996}{701}.}
\lref\sdfp{N. Seiberg, ``Non-trivial Fixed Points of The
Renormalization Group in Six Dimensions,'' hep-th/9609161.}
\lref\Betal{M. Berkooz, R.G. Leigh, J. Polchinski, J. Schwarz,
N. Seiberg, and E. Witten, ``Anomalies, Dualities, and Topology of
$D=6$ $N=1$ Superstring Vacua,'' hep-th/9605184, \np{475}{1996}{115}.}
\lref\Sanom{J. H. Schwarz, ``Anomaly-Free Supersymmetric Models in Six
Dimensions,'' hep-th/9512053, \pl{371}{1996}{223}.}
\lref\aspin{P. S. Aspinwall, ``Point-like Instantons and the
$Spin(32)/Z_2$ Heterotic String,'' hep-th/9612108.}
\def\dim{{\rm dim}}
\def\mH{{\cal M}_H}
\def\mI{{\cal M}_{Inst}}
\def\cA{{\cal A}}

\def\ev#1{\langle#1\rangle}
\def\FI{Fayet-Iliopoulos}
%  draw box of size #1pt and line thickness #2pt
\def\drawbox#1#2{\hrule height#2pt 
        \hbox{\vrule width#2pt height#1pt \kern#1pt 
              \vrule width#2pt}
              \hrule height#2pt}
% Young tableaux

\def\Fund#1#2{\vcenter{\vbox{\drawbox{#1}{#2}}}}
\def\Asym#1#2{\vcenter{\vbox{\drawbox{#1}{#2}
              \kern-#2pt       % line up boxes
              \drawbox{#1}{#2}}}}
 
\def\fund{\Fund{6.5}{0.4}}
\def\asym{\Asym{6.5}{0.4}}
\Title{hep-th/9702038, IASSNS-HEP-97/7}
{\vbox{\centerline{RG Fixed Points in Six Dimensions via}
\centerline{Branes at Orbifold Singularities}}}
\medskip
\centerline{Kenneth Intriligator\footnote{${}^*$}{On leave 1996-1997
{}from Department of Physics, University of California, San Diego.}}
\vglue .5cm
\centerline{School of Natural Sciences}
\centerline{Institute for Advanced Study}
\centerline{Princeton, NJ 08540, USA}

\bigskip
\noindent

We discuss a set of generalized, necessary conditions for non-trivial,
interacting fixed points in six dimensional supersymmetric field
theories.  We use string theory to argue for the existence of infinite
families of interacting RG fixed point theories.  The theories are
based on certain gauge groups and matter content, which we identify,
along with additional tensor multiplets.  They are conjectured to
arise in the world-volume of Type I D5 branes at orbifold
singularities.

\Date{2/97}                                   
%\draftmode                     
                            
\newsec{Introduction}   
                                           
D branes have quantum field theories living in their world volume; see
\jptasi\ for a recent review with references.
For example, in the type I theory, a $SO(32)$ instanton of zero size,
which is a D5 brane, has an $Sp(1)\cong SU(2)$ gauge theory with
$N_f=16$ fundamentals living in its world volume \wsmall.  This 6d
gauge theory is free at long distances and, in the ultraviolet, more
data, such as string theory, is needed to obtain a sensible
theory\foot{I would like to thank N. Seiberg and E. Witten for
instructive comments on this point.}.  The $E_8$ small instanton also
leads to a six dimensional world-volume theory in which gravity
decouples at long distance.  This case appears more exotic -- like a
``tensionless string'' \refs{\gh - \wsd} -- though there is evidence
that it is an interacting {\it local quantum field theory} at a
non-trivial RG fixed point
\refs{\sw, \wsd -\gms}.    
The existence of other non-trivial fixed points of the renormalization
group in six dimensions was discussed in
\sdfp.  For example, it was there argued via string theory that
$SU(2)$ with $N_f=4$ and $N_f=10$ fundamental flavors, along with a
tensor multiplet to cancel gauge anomalies, have non-trivial fixed
points.  It was further conjectured that such fixed points actually
exist for all $N_f<16$ \sdfp. It remains a challenge to formulate and
eventually understand non-trivial six dimensional fixed points
directly in terms of quantum field theory, without having to appeal to
stringy considerations.

Here we will construct new, six-dimensional fixed point theories by
considering type I D5 branes when the Dirichlet location $\vec x\in
R^4$, namely the location of the small $SO(32)$ instantons, is on top
of an orbifold singularity.  We present evidence that, much as with
the small $E_8$ instanton \sw, there is a ``tensionless string'' type
transition region where a ``Higgs branch,'' associated with the
instanton moduli, joins to a ``Coulomb branch,'' associated with
massless tensor multiplets whose scalar components can get expectation
values.  The theories on the Coulomb branch are 6d gauge theories,
with gauge groups and matter content which we identify, along with the
extra tensor multiplets.  We argue that these gauge theories with
extra tensor multiplets have non-trivial RG fixed points at the origin
of their moduli space.

The simplest example of the ``transition'' {}from the Higgs branch to
the Coulomb branch occurs for four D5 branes at a $Z_2$ orbifold
singularity.  With fewer than four instantons, there is no Coulomb
branch. On the Higgs branch, the theory appears to be ``exotic.''
Along the Coulomb branch, on the other hand, the theory is simply
given by an $Sp(4)$ gauge theory with $16$ fundamentals and a tensor
multiplet.  We can approach the RG fixed point at the origin, where
the Higgs and Coulomb branches touch, {}from along the Coulomb
branch. Precisely this ``transition'' {}from the Higgs to the Coulomb
branch also appeared recently in \aspin, where the transition was
discussed in detail in terms of F theory.

In the next section, we will discuss a generalized version of the
anomaly condition of \sdfp\ for non-trivial RG fixed points in six
dimensions.  In sect. 3, we review aspects of small $SO(N)$
instantons.  In sect. 4, we discuss small $SO(N)$ instantons on a
$Z_2$ orbifold singularity.  Following the discussion in \Betal, there
are two situations: with and without ``vector structure\foot{This
terminology is actually being misused here to refer, as discussed
later, to the behavior of a flat connection at infinity.  However, it
was shown in \aspin\ that there can be a non-trivial, ``hidden
obstruction'' $\tilde w_2$ to vector structure even when the flat
connection at infinity is trivial.  In fact, on the Coulomb branch,
the question of vector structure is ill-defined: the obstruction
$\tilde w_2$ can change from being zero to non-zero
\aspin.  For lack of better terminology, we will here 
continue to use ``vector structure'' in a sloppy sense.  I am grateful
to P.S. Aspinwall for a helpful correspondence on this issue.}.''  We
first discuss the case without vector structure, which is simpler; the
essential features in this case already appeared in \refs{\GP,
\Betal}.    In sect. 4.2, we consider the case {\it with} 
vector structure and present evidence that there is a Coulomb branch
associated with the gauge group $Sp(K)\times Sp (K+n-4)$ with $16-2n$
matter fields in the $(\fund , {\bf 1})$, $2n$ matter fields in the
$({\bf 1}, \fund )$, one matter field in the $(\fund , \fund )$, and a
tensor multiplet.  Here $n$ is an arbitrary integer between $0$ and
$8$ and $K$ is an arbitrary integer with $K+n\geq 4$.  One of the
gauge groups, say $Sp(K+n-4)$, is IR free, while the other flows to an
interacting fixed point.  At the fixed point, the theory is $Sp(K)$
with matter $(2K+8)\cdot \fund$ and a tensor multiplet.  String theory
proves that these fixed points really exist for all $K$.  In sect. 5,
we discuss the situation for small $SO(N)$ instantons on a $Z_{M}$
orbifold singularity, with vector structure, for arbitrary $M$.  We
thus argue for the existence of two infinite families of non-trivial
six dimensional fixed points involving gauge theories with extra
tensor multiplets.  In sect. 6 we discuss the case without vector
structure.  The gauge groups and matter hypermultiplets of all of
these theories are given by the type I ``quiver diagrams'' of \dm\
with specific choices of quiver data.

\newsec{Six Dimensional Gauge Theories and Anomalies}

We will be interested in six dimensional theories with the minimal
amount of supersymmetry.  As discussed in \sdfp, there can be
interacting renormalization group fixed points which are related to
gauge theories; a necessary condition for such a fixed point is that
the quartic gauge anomaly can be cancelled.  Here we will discuss the
straightforward generalization of the conditions in \sdfp\ to
non-simple gauge groups with more than one tensor multiplet.

The six dimensional gauge anomaly for a product gauge group 
$G=\prod _aG_a$ is 
\eqn\Agenis{\cA =\sum _a (\Tr F_a^4 -\sum _in_{i}\tr _i F_a^4) -6\sum
_{a<b}\sum _{i<j}n_{ij}\tr _i F_a^2\tr _j F_b^2.} Here $\Tr$ is in the
adjoint representation, $\tr _i$ is in representation $R_i$ of $G_a$,
$n_i$ is the number of matter fields in representation $R_i$, and
$n_{ij}$ is the number of matter fields in representation $(R_i,R_j)$
of $G_a\times G_b$.  This is as in \Sanom, but with the gravitational
parts dropped and a rescaling.  The anomaly \Agenis\ will be of the
form \eqn\Acoeffs{\cA =\sum _a \alpha _a\tr F_a^4+\sum _{ab}c_{ab}(\tr
F_a^2)(\tr F_b^2),} with $\tr$ in the fundamental representation.
Formulae which will be of later use in this regard are:
\eqn\treqs{\matrix{&SU(n):\quad &\Tr F^4=2n\tr F^4+6(\tr F_2)^2,\quad &
\tr _{\asym}
F^4=(n-8)\tr F^4+3(\tr F^2)^2;\cr 
&Sp(n):\quad &\Tr F^4=(2n+8)\tr F^4+3(\tr F^2)^2,\quad &
\tr _{\asym} F^4=(2n-8)\tr
F^4 +3(\tr F^2)^2.}}

If $\cA=0$, the theory is anomaly free and exists with a coupling
constant parameter $g$.  If $\alpha _a=0$ but $c_{ab}\neq 0$ in
\Acoeffs, there is an anomaly which can perhaps be cancelled by
including tensor multiplets.  Much as in \sag, though without
including gravity, the condition for being able to cancel the anomaly
in a theory with $P$ tensor multiplets is
\eqn\Isum{\cA=\sum _{i=1}^P(\sum _a C_{ia}\tr F_a^2)^2}
for some {\it real} constants $C_{ia}$.  In particular, for $G$ simple
and $P=1$ the condition on \Acoeffs\ is $\alpha =0$ and $c>0$ \sdfp.
The anomaly cancellation occurs by coupling the scalar fields $\Phi
_i$ in the tensor multiplets to the gauge fields by the interactions
\eqn\inter{\sum _{i=1}^P\sum _a C_{ia}\Phi _i \tr F_a^2;}
this leads to effective gauge couplings $g_a^{-2}(\Phi
)=g_{a,cl}^{-2}+\sum _iC_{ia} \Phi _i$.

When $\alpha _a=0$ but \Isum\ is not satisfied, the anomalies can
possibly be cancelled by including gravity, as in \Sanom.  If any
$\alpha _a\neq 0$, the theory is inconsistent and can not be cured by
adding any more fields -- even gravity.

\newsec{Small Instantons and the hyper-Kahler quotient}

As discussed in \wsmall, there is an interesting mathematical fact
with a beautiful connection to physics.  The mathematical fact is that
the moduli space of $K$ $SO(N)$ instantons, which is a hyper-Kahler
space, has a ``hyper-Kahler quotient'' construction as the Higgs
branch of a gauge theory with 8 super-charges (i.e. $N=1$ supersymmetry
in 6d or 5d, $N=2$ in 4d, $N=4$ in 3d, etc.)\foot{This can be
generalized to the moduli space of $U(N)$ and $Sp(N)$ instantons, with
the moduli spaces described as the Higgs branches of $U(K)$ and
$SO(K)$ gauge theories, respectively.  The generalization to the
exceptional gauge groups, and in particular $E_8$, appears
problematic; this is connected with the fact that the physics of the
small $E_8$ instanton is exotic.}.  The gauge group is $Sp(K)$ and the
matter content consists of $N$ half-hypermultiplets in the $\fund$
and a hypermultiplet in the $\asym$.

The beautiful connection to physics is that this gauge theory, with
$N=32$, is physically realized: it lives on the world volume of $K$
type I D5 branes at the same point ${\bf x}\in R^4$ fixed by the
Dirichlet boundary conditions.  The anomaly $\cA$ vanishes for any
choice of $K$ \Sanom.  The six dimensional $Sp(K)$ theory exists
as a low energy theory, with coupling constant $g$, which is IR free.
In the UV more data, such as string theory, is needed to obtain a
sensible theory.

\newsec{D5 branes at a $Z_2$ orbifold singularity}

Consider $SO(N)$ instantons on a $Z_2$ orbifold singularity.  The
physical data is specified by an integer $K$, related to the number of
instantons, as well as the possibility of having a non-trivial flat
connection at infinity.  Corresponding to the flat connection at
infinity is an $SO(N)$ group element $\rho_{\infty}$ which has to
satisfy $\rho_{\infty}^2=1$; the possibility of non-trivial
$\rho_{\infty}$ is allowed by the $Z_2$ orbifold identification.
Actually, as discussed in \Betal, there is another possibility in
string theory which makes use of the fact that the gauge group is
really $Spin(32)/Z_2$: we could have $\rho_{\infty}^2=-1$.  As in \Betal,
we will refer to the situation with $\rho_{\infty}^2=1$ as the case {\it
with} vector structure and the situation with $\rho_{\infty}^2=-1$ as
the case {\it without} vector structure.  We will first discuss the
case without vector structure, as it is simpler, with the main points
already appearing in \refs{\GP, \Betal}.

\subsec{The case without vector structure}

Consider generally $Spin (N)/Z_2$, for even $N$, when $\rho_{\infty}$
has $\half N$ eigenvalues $i$ and $-i$.  The second Chern class of the
gauge connection, i.e. the instanton number, is given by
\eqn\viis{I=K+{N\over 32},}
where $K$ is an integer and the last term is a contribution coming
{}from the non-trivial connection at infinity, as explained in \Betal.
The moduli space of instantons with this $\rho_{\infty}$, which is a
hyper-Kahler space, has dimension in hypermultiplet
(i.e. quaternionic) units (one quarter the real dimension) given by an
index theorem to be
\eqn\dimmIv{\dim (\mI )=(N-2)I-{N(N-2)\over 32}=(N-2)K,}
where $-N(N-2)/32$ is the contribution {}from the $\eta$ invariant,
which was discussed for the physical case $N=32$ in \Betal.

The moduli space $\mI$ is conjectured to be isomorphic to the Higgs
branch of a $U(2K)$ gauge theory with $\half N$ matter fields in the
$\fund$ and two in the $\asym$.  These gauge theories were first
derived in \GP\ via an orientifold construction of type I on $K3$.
Note that $\dim (\mH)=\dim (\mI)$, given by \dimmIv.  Further, when a
\FI\ term $\vec \zeta$ (which is a real triplet under the $SU(2)_R$
associated with the Hyper-Kahler structure) is added for the overall
$U(1)$ factor in $U(2K)$, $\mH
\cong \mI$ for instantons on the orbifold with blowing up parameter
$\vec\zeta$, much as in the $U(N)$ instanton case of
\refs{\kn, \dm}\foot{As discussed in some detail in 
\refs{\dm, \Betal}, in 6d
$U(1)$ factors with charged matter are anomalous and become massive
via coupling to a scalar.  The scalar is part of a hypermultiplet
whose expectation value gives the \FI\ term.  In short, the $U(1)$
factors go away and the \FI\ parameter $\vec\zeta$ is replaced with
the expectation value of an additional hypermultiplet.}.

As a check on the conjecture that $\mH\cong \mI$, note that it is easy
to identify the direction in the Higgs branch which corresponds to
moving the $K$ point-like instantons away {}from the orbifold
singularity: give the matter fields in the $\asym$ expectation values
$E_fJ$, where $E_f$ is a hypermultiplet, with $f=1,2$ a flavor index,
and $J\equiv 1_K\otimes (i\sigma _2)$.  This is a $D$-flat direction
provided the $E_f$ satisfy the $D$ term equations associated with the
overall $U(1)$, including the
\FI\ term.  This problem with the $E_f$ is simply that of $U(1)$ with
two electrons, which is precisely the theory whose Higgs branch is
isomorphic to the $Z_2$ orbifold with blowing up parameter $\vec \zeta$ 
(i.e. the $A_1$ ALE or Eguchi-Hanson space) \kron.  The Higgs branch
associated with the $\ev{E_f}$ is thus the modulus for moving the $K$
point-like instantons around on the blown up orbifold.  The point at
the origin, $E_{1,2}=0$, corresponds to the point-like instantons
sitting on the orbifold singularity.  This point is only on the moduli
space when the \FI\ blowing up parameter $\vec \zeta$ is set to zero.
At this point, the full $U(2K)$ is unbroken.  Away {}from this point,
$U(2K)$ is Higgsed to $Sp(K)$ with the correct matter content,
reviewed in the previous section, for describing the $K$ point-like
instantons away {}from the orbifold singularity, or when the singularity
is blown up with $\vec \zeta \neq 0$.

The anomaly \Agenis\ for the $U(2K)$ theory, using \treqs, is $\cA
=(16-\half N)\tr F^4$, and thus vanishes identically for the physical
case of $N=32$.  The theory is free at long distances.

\subsec{The case with vector structure.}
The element $\rho_{\infty}\in SO(N)$ is specified by the number $w _0$
of eigenvalues equal to $+1$ and the number $w_1$ eigenvalues equal to
$-1$, with $w_0+w_1=N$ and $w_1$ even in order to have $\rho_\infty
\in SO(N)$.  To have $\rho _{\infty}^2=1$ in $Spin(N)$, $w_1$ actually
has to be a multiple of four.  With this physical data, the second
Chern class of the gauge connection, i.e. the instanton number, is
given by
\eqn\Iis{I=K+{w_1\over 8},}
where $K$ is an integer and the last term is a contribution coming
{}from the non-trivial connection at infinity, as explained in \Betal.
Note that, $I$ can be half-integral.  The moduli space of such
instantons has dimension in hypermultiplet units given by
\eqn\dimmI{\dim (\mI )=(N-2)I-{1\over 8}w_0w_1;}
the last term is the $\eta$ invariant contribution.

The relevant gauge theory is $Sp(v_0)\times Sp(v_1)$ with $w_0$
half-hypermultiplets in the $(\fund ,{\bf 1})$ representation, $w_1$
half-hypermultiplets in the $({\bf 1}, \fund )$, and one in the
$(\fund , \fund)$.  This corresponds to one of the type I quiver
diagrams of \dm.  Using \treqs, the anomaly
\Agenis\ is
\eqn\Aziigen{\cA=(2v_0+8-\half w_0-2v_1)\tr F_0^4+(2v_1+8-\half
w_1-2v_0)\tr F_1^4+3(\tr F_0^2-\tr F_1^2)^2.}  In general, this theory
has a Higgs branch moduli space of vacua, $\mH$, which is a
hyper-Kahler space, where the hypermultiplets get expectation values.
When there is enough matter to completely Higgs the gauge group, which
will be the case in what follows, the dimension of the Higgs branch is
the number of hyper minus vector multiplets:
\eqn\dimH{\dim (\mH )=\sum _{\mu=0}^1w_\mu v_\mu +4v_0v_1-\sum _{\mu
=0}^1 v_\mu (2v_\mu +1).}

We conjecture that $\mI \cong \mH$ with the following relation between
the $v_\mu$ and the physical data:
\eqn\vkwhk{v_0=K,\quad v_1=K+{w_1\over 4} \qquad \rm{(Hyper-Kahler\
quotient)}.}
Note that, with this identification, the instanton number \Iis\ is
given by
\eqn\Iisv{I=\half (v_0+v_1),}
which is quite natural upon considering the $Sp(v_\mu)$ on the
covering space of the $Z_2$ orbifold.  Also, substituting in
\vkwhk, it is seen that the dimensions in \dimmI\ and
\dimH\ agree.  

As a further check on $\mH \cong \mI$, we note that the gauge theory
properly has moduli which correspond to moving small instantons away
{}from the $Z_2$ orbifold singularity: The $Sp(K)\times Sp(K+{1\over
4}w_1)$ theory has a set of flat directions on the Higgs branch where
the matter field in the $(\fund , \fund )$ gets an expectation value,
breaking the gauge group to $Sp(R) _D\times Sp(K-R)\times
Sp(K-R+{1\over 4}w_1)$, where $Sp(R)_D$ is diagonally embedded in the
two original gauge groups.  This flat direction exists for any
$R=1\dots K$ and corresponds to moving $R$ of the point-like
instantons away {}from the $Z_2$ singularity.  In particular,
$Sp(R)_D$ has exactly the standard matter content of $R$ small
instantons away {}from the $Z_2$ orbifold singularity.

Despite this evidence for $\mH \cong \mI$ with the data \vkwhk, there
is an important caveat which was already apparent in the work of \dm.
In the analogous hyper-Kahler quotient construction of \kn\ for $U(N)$
instantons on ALE spaces, and in the example of the previous
subsection, it was possible to blow up the orbifold singularity via
\FI\ terms in the hyper-Kahler quotient gauge theories.  In the
present situation, however, the $Sp(v_0)\times Sp(v_1)$ gauge theory
does not have any overall $U(1)$ factor, as would be needed to add a
\FI\ term.  Therefore, the hyper-Kahler quotient construction of $\mI$
discussed here only applies when the $Z_2$ orbifold singularity is
{\it not} blown up; the corresponding blowing up mode is missing.
While we will not find this missing mode, it should not be forgotten.
Below we will argue that, in the physical $SO(32)$ string, it nicely
joins with 28 other ``missing'' moduli in a transition where $29$
hypermultiplets are traded in for a tensor multiplet.  

The $Sp(v_0)\times Sp(v_1)$ theory with the data \vkwhk\ can {\it not}
arise as a world volume gauge theory of D5 branes at an orbifold
singularity.  This is because, with the data \vkwhk, the anomaly
\Aziigen\ is non-zero.  In fact, the coefficients of the deadly $\tr
F_\mu ^4$ terms are non-zero, so the theory has a sickness which can
not possibly be cured, even by adding more fields.  There is, though,
a simple modification of the data \vkwhk\ for which the $\tr F_\mu ^4$
terms in \Aziigen\ do drop out:
\eqn\vkwp{v_0=K,\quad v_1=K+{w_1\over 4}-4 \qquad \rm{(Physical)}.}
With this choice, the $\tr F_0^4$ term properly vanishes for $N=32$
and the coefficient of the $\tr F_1^4$ term is identically zero.
Furthermore, the remaining terms in \Aziigen\ are a perfect square and
thus the anomaly can be cancelled by including a single tensor
multiplet.

With the choice of data \vkwp, the Higgs branch of the $Sp(v_0)\times
Sp(v_1)$ gauge theory is no longer isomorphic to the moduli space of
instantons on the $Z_2$ singularity; even their dimensions differ.
Applying the general formulae \dimmI\ and
\dimH, we see that Higgs branch of the theory with
data \vkwp\ has $28$ fewer hypermultiplet moduli than the moduli space
of instantons.  In addition, there is the missing blowing up mode
mentioned above.  All together, we are missing $29$ hypermultiplets
and need an extra tensor multiplet.  

This trade between $29$ hypermultiplets and a tensor multiplet is
familiar in compactifications of string theory to six dimensions,
where it is crucial that both make the same contribution to the purely
gravitational anomaly \refs{\lagw, \dmw}.  The classic example is the
small $E_8$ instanton theory: there is a ``Higgs'' branch, with 29
massless hypermultiplets, which touches a ``Coulomb'' branch, with a
massless tensor multiplet, at a transition point in the moduli space
\sw.  This point is where the $E_8$ instanton is of zero size.  Other
examples of such transitions involving extra tensor multiplets have
been found in a variety of string compactifications to 6d.

The picture that we have for what happens when $K$ D5 branes sit on
the $Z_2$ orbifold singularity is the following: There is an
interesting six dimensional quantum field theory living on the world
volume.  This theory has a moduli space with two branches: ``Higgs''
and ``Coulomb,'' which join near the origin.  The Higgs branch is
isomorphic to the full moduli space of instantons on the orbifold,
including a mode for blowing up the orbifold.  Near the origin, with
28 of the instanton moduli and the blowing-up mode set to zero, the
theory has a ``Coulomb'' branch with a massless tensor multiplet,
whose real scalar component $\Phi$ can get an expectation value.  On
this Coulomb branch is the $Sp(K)\times Sp(K+{1\over 4}w_1-4)$ gauge
theory with a tensor multiplet.  This situation is quite analogous to
that of the small $E_8$ instanton picture, with the $Sp\times Sp$
gauge theory super-imposed on the tensor multiplet of the Coulomb
branch.

Consider for simplicity the case with $w_1=0$.  In order for the above
gauge theory to make sense, $K\geq 4$; for $K<4$ there is no Coulomb
branch.  For $K=4$ small instantons sitting on the $Z_2$ singularity,
there is a Coulomb branch to a theory which is similar to the usual
$Sp(4)$, with matter $16\cdot \fund$, though without the usual $\asym$
in the ${\bf 27}+{\bf 1}$ of $Sp(4)$, and with an extra tensor
multiplet.  The missing ${\bf 27}+{\bf 1}$ means that all four
instantons are locked at the $Z_2$ orbifold point.  In addition, the
mode for blowing up the $Z_2$ orbifold singularity is locked to zero.
These modes get a mass along the Coulomb branch.  In the place of
these $29$ missing hypermultiplets is the extra tensor multiplet.
Exactly this situation was independently found recently by Aspinwall
to occur in compactification of the heterotic string to six dimensions
on $K3$ \aspin, where it was found via $F$ theory.

Along the Coulomb branch, as in the discussion following
\inter, the cancellation of the remaining anomaly in \Aziigen\ means that
\eqn\ziigeff{g^{-2}_{0,eff}=g^{-2}_{0,cl}+\Phi,\qquad
g_{1,eff}^{-2}=g^{-2}_{1,cl}-\Phi,} where we absorbed a normalization
constant into the normalization of $\Phi$.  The theory has a single
coupling constant parameter: $g_0^{-2}+g_1^{-2}$.  This is sensible
because this is precisely the coupling of the $Sp(R)_D$ theory
obtained upon moving $R$ of the point like instantons away {}from the
singularity.  We can set $g^{-2}_{0,cl}=0$ in \ziigeff, by choice of
the origin of $\Phi$, and take the Coulomb branch to be $\ev{\Phi}\in
R^+$.  The $Sp(K)$ theory can have a non-trivial RG fixed point at the
origin of the Coulomb branch.  On the other hand, the coupling
$g_{1,eff}^{-2}$ of $Sp(K+{1\over 4} w_1-4)$ has a ``Landau pole'' on
the Coulomb branch at $\ev{\Phi}\geq g_{1,cl}^{-2}$.  This reflects
the fact that the $Sp(K+{1\over 4} w_1-4)$ theory is free at long
distances and more data, such as string theory, is needed in the
ultraviolet to obtain a sensible theory.  This aspect of the theory is
inherited {}from the theory of small instantons away {}from
singularities, which is also IR free.

The transition point between the Higgs and Coulomb branches is
associated with ``tensionless strings,'' which is interpreted as a
signature of an interacting, local, RG fixed point. As discussed
above, only the $Sp(K)$ gauge group can have a non-trivial fixed
point.  At long distances, the $Sp(K+{1\over 4}w_1-4)$ theory is IR
free and thus becomes un-gauged.  The theory at the fixed point is
then $Sp(K)$ with matter given by $(2K+8)\cdot \fund$, along with a
tensor multiplet.  The string theory construction proves the existence
of RG fixed points at the origin of the Coulomb branch of these
theories.

\newsec{Small Instantons at $Z_M$ orbifold singularities with vector
structure}

We now generalize the analysis of the previous section to arbitrary
$Z_M$ orbifolds with vector structure.  Let $w_\mu$ be the number of
eigenvalues of $\rho_{\infty}$ equal to $e^{2\pi i\mu /M}$ for $\mu
=0\dots M-1$.  Having $\rho_{\infty}\in SO(N)$ requires
$w_{\mu}=w_{M-\mu}$ and $\sum _{\mu =0}^{M-1}w_\mu =N$.  For $M$ even,
having $\rho ^M=1$ in $Spin(N)$ actually requires $w_{{M\over 2}}$ to
be a multiple of four.

The instanton number is given by
\eqn\Iisgen{I=K+\sum _{\mu=0}^{M-1}{\mu(M-\mu)\over 4M}w_\mu,}
where $K$ is an arbitrary integer and the last term is the
contribution to the second Chern class of the non-trivial flat
connection of $\rho_{\infty}$.  The dimension of the moduli space of
instantons with these physical data is given by an index theorem to be
\eqn\dimIgen{\dim (\mI )=(N-2)I+\half (\sum _{\mu, \nu
=0}^{M-1} w_\mu w_\nu X_{\mu \nu}-\sum _{\mu =0}^{M-1}w_{\mu}X_{\mu,
M-\mu}),}
where the last terms are the $\eta$ invariant, with $X_{\mu \nu}$
defined by 
\eqn\Xis{X_{\mu \nu}\equiv
{1\over 2M}\sum _{k=1}^{M-1}{
e^{2\pi i k(\mu-\nu)/M}-1\over 2-2\cos({2\pi
k\over M})}= -{|\mu -\nu |(M-|\mu -\nu |)\over 4M}.} 

The relevant gauge theory depends on whether $M$ is even or odd; in
either case we write $M=2P$ or $M=2P+1$.  The two cases are given by,
respectively, the ``I.4'' and ``I.2'' quiver diagrams described in
sect. 4.4 of
\dm. For $M=2P$ the gauge theory
is \eqn\evengg{Sp(v_0)\times U(v_1)\times U(v_2)\times \cdots \times
U(v_{P-1})\times Sp (v_P), \qquad M=2P, } with hypermultiplets $\half
w_0 \cdot \fund _0$, $\oplus _{j=1}^{P-1}w_j\cdot \fund _j$, $\half
w_P\cdot \fund _P$, and $\oplus _{j=1}^{P}(\fund _{j-1} , \fund _j)$
(subscripts label the gauge group).  For $M=2P+1$ the relevant gauge
theory is
\eqn\oddgg{Sp(v_0)\times U(v_1)\times U(v_2)\times \cdots \times
U(v_{P-1})\times U(v_P), \qquad M=2P+1, } with hypermultiplets in the
$\half w_0 \cdot \fund _0$, $\oplus _{j=1}^{P}w_j\cdot \fund _j$,
$\oplus _{j=1}^P (\fund _{j-1} , \fund _j)$, and $\asym _P$.  It will
be convenient in what follows to define $V_{\mu}$ for $\mu=0
\dots M-1$ by $V_{\mu\leq P}\equiv \dim (\fund _\mu)$ and $V_{\mu
>P}\equiv V_{M-\mu}$; i.e. $V_0\equiv 2v_0$, $V_{i<P}\equiv v_i$, and
$V_P\equiv 2v_P$ ($V_P\equiv v_P$) for $M=2P$ ($M=2P+1$).

The anomaly \Agenis\ for the above theories is given, using \treqs, by
\eqn\AzMis{\cA =\half \sum _{\mu =0}^{M-1}(
\widetilde C_{\mu \nu}V_\nu -w_\mu +D_\mu)\tr F_\mu ^4+3\sum
_{j=1}^P(\tr F_{j-1}^2 -\tr F_j^2)^2;} $F_\mu$ for $\mu >P$ is defined
by $F_\mu \equiv F_{M-\mu}$, $\widetilde C_{\mu\nu}\equiv 2\delta
_{\mu \nu}-a_{\mu \nu}$ is the Cartan matrix of the extended $SU(M)$
Dynkin diagram, and $D_\mu \equiv 8(2\delta _{\mu, 0}+\delta _{\mu
,P}+\delta _{\mu ,M-P})$.  We will also be interested in the Higgs
branch; when there is sufficient matter to completely Higgs the gauge
group, as will be the case below, its dimension can be written as
\eqn\dimHM{\dim (\mH )=\half V_\mu w_\mu -{1\over
4}\widetilde C_{\mu \nu}V_\mu V_\nu -\half (V_0+V_P).} Throughout,
repeated greek indices are summed {}from $0$ to $M-1$.

The above $V_{\mu}$ will be related to the $w_\mu$ and $K$ by $V_0=2K$
and $\widetilde C_{\mu \nu}V_\nu =w_\mu -u_\mu$ for some $u_\mu$
satisfying $u_\mu n_\mu=w_\mu n_\mu =N$, with $n_\mu \equiv 1$ for all
$\mu$.  Thus,
\eqn\Vwu{V_{0}=2K, \qquad V_{i\neq 0}=2K+\sum
_{j=1}^{M-1}C^{-1}_{ij}(w_j-u_j),} with $C^{-1}_{ij}$ the inverse
$SU(M)$ Cartan matrix, given by $C^{-1}_{i<j}=i(M-j)/M$.  With \Vwu,
\eqn\Vsum{{1\over M}\sum _{\mu =0}^Pv_{\mu}={1\over 2M}v_\mu n_\mu 
=K+\sum _{\mu =0}^{M-1}{\mu (M-\mu)\over 4M}(w_\mu -u_\mu)=I-\sum
_{\mu =0}^{M-1} {\mu (M-\mu)\over 4M}u_\mu,} Also, It follows {}from
\Vwu\ and some straightforward but tedious manipulations that
\eqn\dimih{\dim (\mI )=\dim (\mH)-\sum _{j=1}^{M-1}((u_\nu n_\nu
-u_0)X_{0,j}+\half C^{-1}_{P,j})u_j+\half
\sum _{i,j=1}^{M-1}u_iu_j X_{ij}.}

We conjecture that $\mH \cong \mI$ for $u_\mu =N\delta _{\mu ,0}$ in
\Vwu.  Explicitly, 
\eqn\VMhk{V_{\mu \leq P} =2K+\sum _{\nu =0}^P{\rm min}(\mu ,
\nu)W_\nu \qquad ({\rm Hyper-Kahler\ quotient}),}
where $W_{\mu <P}\equiv w_\mu$ and $W_P\equiv w_P$ ($W_P\equiv \half
w_P$) for $M$ odd (even).  In this case, \Vsum\ is simply the
instanton number, which is quite natural in terms of considering the
instantons on the $M$-fold cover of the orbifold.  In addition, in
this case the gauge group
\evengg\ can be completely Higgsed and it follows {}from \dimih\ that
$\dim (\mH ) =\dim (\mI )$.

Note that the gauge groups \evengg\ and \oddgg\ have $M-P-1$ overall
$U(1)$ factors to which \FI\ terms can be coupled.  Adding such terms
corresponds to turning on blowing-up moduli of the orbifold.  However,
because the $Z_M$ orbifold should have $M-1$ blowing up moduli, we see
that $P$ of the blowing up modes are locked at zero.  As in the
previous section, these $P$ missing modes will be remembered below.

The theory with data \Vwu\ for $u_\mu =N\delta _{\mu 0}$, however, can
{\it not} arise in the world volume of D5 branes on the orbifold
singularity.  This is because this theory has a deadly $\tr F_P^4$
anomaly term with non-zero coefficient.  There is an obvious
modification of \Vwu, though, for which all $\tr F_\mu^4$ terms do
\AzMis\ vanish: $u_\mu =D_\mu \equiv 8(2\delta _{\mu 0}+\delta _{\mu
P}+\delta _{\mu ,M-P})$; which requires $N=32$, as expected.  More
explicitly, with this modification \VMhk\ becomes
\eqn\VMp{V_{\mu \leq P} =2K+\sum _{\nu =0}^P{\rm min}(\mu ,
\nu)W_\nu -8\mu \qquad ({\rm Physical}),}
With this choice
of data, \AzMis\ becomes
\eqn\AzMn{\cA =3\sum
_{j=1}^P(\tr F_{j-1}^2 -\tr F_j^2)^2,} which can be nicely cancelled
by coupling the theory to $P$ tensor multiplets.

With the choice of data \VMp, \Vsum\ gives 
\eqn\vMtott{{1\over M}\sum _{\mu =0}^Pv_{\mu}=I-(M-{M-2P\over M}).}
This difference has a natural interpretation: the $v_\mu$ make up the
difference between the instanton number and that of the standard
embedding.  With the standard embedding, one should have
$I_{s.e.}=\chi$, the Euler character.  Actually, there are two
possible choices for $\chi$ associated with the $Z_{M}$ orbifold: that
of the bulk, which is $\chi _{bulk}=M-M^{-1}$, and one which includes
a contribution associated with the boundary at infinity, $\chi
_{total}=
\chi _{bulk} +\chi _{boundary}=M$.  For $M=2P+1$, the $v_i$ make up
the difference between the instanton number $I$ and $I_{s.e.}=\chi
_{bulk}$.  For $M=2P$ the $v_i$ make up the difference between $I$ and
$I_{s.e.}=\chi _{total}$.  The fact that all $v_\mu =0$ when
$I=\chi _{bulk}$ for $M=2P+1$ is related to the result that there are
no five-branes in compactification of type I on a $Z_3$ orientifold
$K3$ \GJ.

In addition, with \VMp\ it is no longer true that $\mH \cong
\mI$.  Indeed, even their dimensions differ: as seen {}from \dimih, 
\eqn\diffdim{\dim (\mH )=\dim (\mI)-28P;}
we are missing $28P$ hypermultiplet moduli.  Combining these with the
$P$ missing blowing-up modes mentioned above, we see that a transition
has occurred where $29P$ hypermultiplets have been traded for $P$
tensor multiplets.  This is perfect, because $P$ tensor multiplets is
precisely what we needed above to cancel the anomaly \AzMn.

We thus conjecture that the theories \evengg\ or \oddgg\ with the data
\VMp, and $P$ tensor
multiplets, arise as a ``Coulomb branch'' of small instantons on
orbifold singularities with vector structure.  Note that the Coulomb
branch can only exist when all \VMp\ satisfy $V_{\mu}\geq 0$.  In
particular, for $\rho _{\infty}=1$ (i.e. $w_\mu =32\delta _{\mu ,0}$),
the condition for there to be a Coulomb branch is $K\geq 4P$.

Cancelling \AzMn\ via
\inter, the effective gauge couplings on the Coulomb branch satisfy
\eqn\geffcg{g_{i,eff}^{-2}=g^{-2}_{i,cl}+\Phi_{i+1}-\Phi_i, \qquad
i=0\dots P,} where $\Phi _0\equiv \Phi _{P+1}\equiv 0$ and
normalization constants are absorbed into the normalization of $\Phi
_i$.  There is thus a single parameter coupling, $\sum
_{i=0}^Pg_i^{-2}$; this is the coupling of the $Sp(R)$ theory,
obtained along a flat direction, which is associated with moving $R$
instantons away {}from the singularity.  By a choice of the origin of
the $\Phi _i$, it is possible to take $g_{i<P,cl}^{-2}=0$ in
\geffcg.  The Coulomb branch is then given by the {\it wedge} in
$R^P$, $0\leq \Phi _1\leq \Phi _2 \dots \leq \Phi _P$, with
$g^{-2}_{i<P,eff}\geq 0$ in the entire wedge.  On the other hand,
$g_{P,eff}^{-2}$ hits a Landau pole when $\ev{\Phi _P}\geq
g_{P,cl}^{-2}$.  This is a reflection of the fact that the last
$Sp(v_P)$ factor in
\evengg\ or $U(v_P)$ factor in \oddgg\ is IR free and UV incomplete.  
This property is inherited {}from the theory discussed in sect. 3.

At the origin of the Coulomb branch, $\ev{\Phi _i}=0$, there is an
interacting RG fixed point associated with \evengg\ or \oddgg\ and the
data \VMp.  However, because the last
$Sp(v_P)$ or $U(v_P)$ factor in \evengg\ or \oddgg\ is IR free, this
factor is not gauged at the fixed point obtained at long distances.
Summarizing, then, the fixed point theories have gauge group
$Sp(v_0)\times \prod _{i=1}^{P-1}U(v_i)$, with data \VMp,
hypermultiplets $\half w_0\cdot \fund _0$, $\oplus
_{j=1}^{P-1}(w_j+V_P\delta _{j,P-1})\cdot \fund _j$, $\oplus
_{j=1}^{P-1}(\fund _{j-1}, \fund _j)$, and $P$ tensor multiplets.  The
string theory construction in terms of branes at orbifold
singularities shows that these fixed points really exist.

\newsec{Small Instantons at $Z_{2P}$ orbifold singularities 
without vector structure} We now briefly turn to the cases without
vector structure.  For $M=2P$, $\rho _\infty$ has $w_i$ eigenvalues
$e^{i\pi (2i-1)/2P}$, $i=1\dots 2P$, with $w_{2P+1-i}=w_i$ and
$\sum _{i=1}^{2P}w_i=2\sum _{i=1}^P=N$ for $\rho _{\infty}\in SO(N)$.
It is expected that the relevant gauge theory is that of the ``type
I5'' quiver diagrams of \dm.  The gauge group is $\prod
_{i=1}^PU(v_i)$ and the matter content is $\oplus _{i=1}^Pw_i\cdot
\fund _i$, $\oplus _{i=1}^{P-1}(\fund _i, \fund _{i+1})$, $\asym _1$,
and $\asym _P$.  Note that, as there are only $P$ $U(1)$ factors to
which \FI\ terms can be coupled, there are $P-1$ missing blowing up
modes.

The anomaly \Agenis\ is found to be
\eqn\anov{\cA=\half \sum _{i=1}^{2P}(\sum _{j=1}^{2P}
\widetilde C_{ij}v_j-w_i+D_i)\tr F_i^4+3\sum _{r=1}^{P-1}(\tr F_r^2
-\tr F_{r+1}^2)^2,} where $\widetilde C _{ij}$ is the Cartan matrix
for the extended $SU(2P)$ Dynkin diagram, $F_{i>P}\equiv F_{2P+1-i}$,
$v_{i>P}\equiv w_{2P+1-i}$, and $D_i\equiv 8(\delta _{i,1}+\delta
_{i,P}+\delta _{i,2P}+\delta _{i, P+1})$.  These theories thus have an
anomaly which can be cancelled with $P-1$ tensor multiplets provided
\eqn\vwnovec{\sum _{j=1}^{2P}\widetilde C_{ij}v_j=w_i-D_i.} Note that
this properly gives $N=32$.

Cancelling \anov\ via \inter, the effective gauge couplings are,
\eqn\geffnov{g_{i,eff}^{-2}=g_{i,cl}^{-2}+\Phi _{i}-\Phi _{i-1}, \qquad 
i=1\dots P,} with $\Phi _0\equiv \Phi _P\equiv 0$ and normalization
factors absorbed into the normalization of the $\Phi _i$.  There is
thus a single coupling parameter, $\sum _{i=1}^P g_i^{-2}$, which is
that of the $Sp(R)$ gauge theory of small instantons away {}from the
singularity.  By choice of the origin of the $\Phi _i$, we can take
$g_{i<P,cl}=0$ in \geffnov\ and the Coulomb branch to be given by the
wedge in ${\bf R}^{P-1}$: $0\leq \Phi _1\leq
\Phi _2\cdots \leq \Phi _{P-1}$.  The $g^{-2}_{i<P,eff}\geq 0$
everywhere on the Coulomb wedge.  On the other hand, $g^{-2}_{P,eff}$
hits a Landau pole at $\ev{\Phi _P}\geq g_{P,cl}^{-2}$.  This
corresponds to the fact that the last $U(v_P)$ factor is IR free.

We conjecture that these theories arise as the Coulomb branch of
the world volume theory of D5 branes on orbifold singularities without
vector structure.   Some examples were
explicitly constructed via orientifold techniques in
\refs{\GJ, \DP}: the ``$Z_4^A$'' orientifold 
there gave the $P=2$ case, with $v_i=w_i=8$, and the ``$Z_6^A$''
orientifold gave the $P=3$ case, with $v_i=w_i$, $w_1=w_3=4$, $w_2=8$.
These restrictions came {}from compactifying on a compact $K3$.  We
conjecture that all 6d gauge theories of the above type satisfying
\vwnovec\ can be obtained in the world volume of D5 branes at
(non-compact) orbifold singularities without vector structure.

The $\prod _{i=1}^{P-1}U(v_i)$ part of the above theories have
non-trivial RG fixed points at the origin of the Coulomb branch.  The
$U(v_P)$ part is IR free and thus un-gauged at long distances\foot{In
all of the examples, by changing the sign of the range of the
$\ev{\Phi _i}$ on the Coulomb branch, the first rather than the last
gauge group can be taken to be the one which is IR free.  For the
gauge theories \evengg\ and those discussed in this section, this
operation simply leads to the same fixed point theories.  On the other
hand, the fixed point theories described in this section could,
alternatively, have been obtained {}from \oddgg\ via this operation.}.
The string theory construction shows that all of these fixed points
really exist.

\bigskip
\centerline{{\bf Acknowledgments}}

I would like to thank J. Blum, M. Douglas, N. Seiberg, and E. Witten
for useful discussions.  The work of K.I. is supported by NSF
PHY-9513835, the W.M. Keck Foundation, an Alfred Sloan Foundation
Fellowship, and the generosity of Martin and Helen Chooljian.
\listrefs
\end